
\magnification=1200

\parskip 3pt plus 1pt minus 1pt
\rightline{DTP/93/09}
\rightline{March, 1993
}
\vskip 2 true cm
\centerline{\bf{STUDY OF QUOMMUTATORS}}
\vskip 0.5 true cm
\centerline{\bf{OF QUANTUM VARIABLES AND GENERALIZED DERIVATIVES.}}
\vskip 3 true cm
\centerline{\bf{David FAIRLIE}}
\vskip 0.5 true cm
\centerline{Department of Mathematical Sciences}
\centerline{University of Durham, Durham, DH1 3LE, England}
\vskip 1 true cm
\centerline{\bf{Jean NUYTS}}
\vskip 0.5 true cm
\centerline{University of Mons, 7000, Mons, Belgium}
\vskip 5 true cm
\noindent Abstract :
\smallskip
A general deformation of the Heisenberg algebra is introduced with two deformed
operators instead of just one. This is generalised to many variables, and
 permits the simultaneous existence of coherent states, and the transposition
of
 creation operators.

\vfill\eject
{\bf{I. Introduction}}
\vskip 0.5 true cm
\par Let $X_i$ $(i=1,\ldots,N)$ be a set of quantum variables which obey the
 quommutation relations
$$
X_i*X_j=x_{ij} X_j*X_i
\eqno(1.1a)
$$
where the c-numbers $x_{ij}$ obviously satisfy
$$
x_{ij}x_{ji}=1
\eqno(1.1b)
$$
and $*$ represents an associative non commutative product. This is not by any
 means the most general starting point$^{[1]}$, but is the most convenient
 choice
as associativity is automatically satisfied for the $X_i$.
Let $D_i$ $(i=1,\ldots,N)$ the corresponding quantum derivatives
$$
D_i*D_j=d_{ij} D_j*D_i
\eqno(1.2a)
$$where again
$$
d_{ij}d_{ji}=1\ \ \ .
\eqno(1.2b)
$$
\par It is then
useful to introduce the operators $G_i$ $(i=1,\ldots,N)$ defined for each $i$
by
$$
G_i=X_i*D_i
\eqno(1.3)
$$
which as a consequence of (1.1) and (1.2) satisfy similar quommutation
relations
$$
G_i*G_j=g_{ij} G_j*G_i
\eqno(1.4a)
$$
and
$$
g_{ij}g_{ji}=1\ \ \ .
\eqno(1.4b)
$$
The operators $G_i$ are to be interpreted as quantum dilatation operators
analogous to the classical dilatation operator usually defined as
$X_i{\partial / \partial_{X_i}}$.
\par Since we expect the quommutator of the operator $D_i$ with the
corresponding
$X_i$ to involve diagonal neutral operators$^{[2]}$ which we call
$A_i$ $(i=1,\ldots,N)$, we write
$$
D_i*X_j=v_{ij} X_j*D_i+\delta_{ij}A_i
\eqno(1.5)
$$
where the arbitrary normalisation of the diagonal $\delta_{ij}$ term has been
supposed to
be non zero for all $i$ and has been included in the definition of the
$A_i$.
\par The motivation for this study, is connected with the potentiality of
representing $A$ by something other than the identity operator. This will
permit us to overcome a deficiency which has been
ignored by much of the recent
literature on q-deformations\rlap,$^{[3,4]}$ and will enable us to
demonstrate
the existence of coherent states and at the same time permit transposition of
the variables $X_i$. As far as we are aware, there are only certain
particularly simple parameter choices in some recent work on differential
calculus$^{[4,5]}$ which currently allow this.
A secondary motivation is to write an operator realisation of  the symmetric
 form of the q-derivative, and generalise this to many variables.
\par We can write after multiplication of (1.5) on the left by
$X_i$
$$
G_i*X_j=p_{ij} X_j*G_i+\delta_{ij}X_i*A_i
\eqno(1.6a)
$$
where
$$
p_{ij}=v_{ij}x_{ij}
\eqno(1.6b)
$$
\par Let us define naive scale invariance as follows : $X_i$ scales as
a length $[L^1]$, $D_i$ as an inverse length $[L^{-1}]$ and hence $G_i$
and $A_i$ as $[L^0]$. This naive scale invariance would allow the
addition of a term
$\rho_iX_i$  to
the right hand side of (1.6a). We have however shown that the associativity
restrictions on the operators lead to $\rho_i=0$ in a natural way (except
for pathological configurations which we have excluded).
\par In order to construct a complete set of quommutator relations the
equations (1.1),
(1.4) and (1.6) have to be supplemented by relations between the $A_i$
themselves as well as between the $A_i$ and both the $X_j$ and the $G_j$.
\par For these quommutation relations we propose
$$\eqalignno{
A_i*A_j&=a_{ij} A_j*A_i      &(1.7a)\cr
a_{ij}a_{ji}&=1              &(1.7b)\cr}
$$
i.e. of a form analogous to (1.1), (1.4) and
$$\eqalignno{
A_i*X_j&=q_{ij} X_j*A_i+\delta_{ij}\alpha_i X_i*G_i &(1.8) \cr
G_i*A_j&=r_{ij} A_j*G_i+\delta_{ij}(\beta_i A_i^2
                        +\gamma_i G_i^2)  &(1.9) \cr}
$$
\par Let us state again that in (1.8) we could have added a term $\sigma_iX_i$
and in (1.9) a term $\lambda_iA_i+\mu_iG_i+\nu_i$ in agreement with the
naive scale invariance and the idea that all terms with degree less or
equal to two should be included in these relations. But all these terms
turn out to be zero $\sigma_i=\lambda_i=\mu_i=\nu_i=0$, in a natural
way, upon the application of the
associativity requirements. Here again it would be possible to generalise, by
 intermixing terms with different indices as is done$^{[5-8]}$
, but we take the
 simplest choice. As we shall see this will permit us to change the basis to
simplify the algebra. What we have is a generalisation of previous
work$^{[9]}$, with the
 incorporation of the $A_i$ operators.
Once the form (1.8--9) is obtained both
$A_i$ and $G_i$ can still be renormalized (rescaled) by the same factor.
\par The rules given above enables one to rewrite any product of operators
in what we shall call a normal order i.e.
\item{-} the $G_i$ at the right of all the other operators ($X_i$ and $A_i$)
\item{-} the $A_i$ at the right of the $X_i$
\item{-} within the operators of the same name, the operators are ordered in,
say, decreasing order of their indices.
\par There is a however a subtle point connected to (1.9). If one takes
that equation for $i=j$, i.e.
$$
G_i*A_i=r_{ii} A_i*G_i+\beta_i A_i^2 +\gamma_i G_i^2
\eqno(1.10)
$$
and multiplies it by $A_i$ on the right, one obtains, dropping the index
$i$ as in (2.3)
below, using (1.10) repeatedly and after
rearrangement of the terms
$$
G*A^2=\beta\gamma G*A^2+\beta(1+r) A^3+\gamma^2(1+r) G^3
         +r\gamma(1+r)A*G^2+r(r+\beta\gamma)A^2*G
\eqno(1.11)
$$
All the terms in the right hand side, except the first one are normal
ordered. There is thus a condition to be able to write $G*A^2$ in normal
order : namely that
$$
\beta\gamma \neq 1
\eqno(1.12)
$$
which we shall call the ``normal ordering condition''.
In this case
$$\eqalign{
G*A^2={1\over (1-\beta\gamma)}
              &{\big(} \beta(1+r) A^3+\gamma^2(1+r) G^3\cr
         & +r\gamma(1+r)A*G^2+r(r+\beta\gamma)A^2*G {\big)} \cr}
\eqno(1.13)
$$
\par The condition (1.12), essential to the consistency of the definition of
the
normal product, will be systematically imposed in the next section.
\par In section 2 we solve, in general, the associativity requirements
which, as we know, are not automatic for quommutation relations of the form
given above. In section 3 we outline definitions of symmetric quantum
derivatives which give representations of one of the solutions obtained
in section 2.
\vskip 0.5 true cm
{\bf{II. Solutions of the associativity requirements.}}
\vskip 0.5 true cm

\par We present here the general restrictions on the parameters which
follow from the braiding relations which  ensure that
any product of operators can be rewritten in an unambiguous unique way as a
normal product. Let us note that we have excluded
pathological solutions such that $x_{ij}=0$, $p_{ij}=0$, $\ldots$.
\par After some lengthy computations one finds that associativity
for the products
$$\eqalign{
A*X*X&,\ \ \ \ G*X*X,\cr
G*A*A&,\ \ \ \ A*A*X,\cr
G*G*X&,\ \ \ \ G*G*A \cr}
\eqno(2.1)
$$
for all the indices $i,j,k$ require
very simple equalities between the non-diagonal coefficients
$$\eqalignno{
r_{ij}&=a_{ij}=g_{ij}\ \ \ \ \  &(2.2a)   \cr
&\ \ \ \ \ \ \ \ \ \ \ \ \ \ \ \ \ \ \ \ \ \ \ \ \ \ \ \ \ \ i\neq j & \cr
q_{ij}&=p_{ij} \ \ \             &(2.2b)   \cr}
$$
while the coefficients of the other possible terms in the right hand sides,
i.e. the linear $(\lambda,\sigma,\mu,\rho)$ terms as well as the constant
term $\nu$, are forced, for non pathological solutions, to be zero.
\par The treatment of the diagonal case is somewhat more complicated.
\par Indeed the relations imposed by associativity which remain to be checked
are those for
$(G_i*A_i)*X_i=G_i*(A_i*X_i)$, i.e. within one family only. Dropping the
index $i$ one has
to find the restrictions on the parameters for the set of quommutators
$$\eqalignno{
G*X&=p X*G+X*A &(2.3a)  \cr
A*X&=q X*A+\alpha X*G &(2.3b)\cr
G*A&=r A*G+\beta A^2+\gamma G^2 &(2.3c)\cr}
$$
with $p=p_{ii}$, $\alpha=\alpha_i$, $\ldots$.
\par There are four disconnected cases which we now give explicitely
\vskip 0.5 true cm
\item{Case A}
$$\eqalign{
G*X&=p X*G+X*A   \cr
A*X&=q X*A+pq X*G \cr
G*A&=r A*G+\beta A^2+(q-qr-q^2\beta) G^2 \cr}
\eqno(2.4Aa)
$$
with the normal ordering condition (1.12)
$$
\beta(q-qr-q^2\beta)\neq 1
\eqno(2.4Ab)
$$
\vskip 0.5 true cm
\item{Case B}
$$\eqalign{
G*X&=p X*G+X*A   \cr
A*X&=q X*A+\alpha X*G \cr
G*A&=(1+\beta p-\beta q) A*G+\beta A^2-\alpha\beta G^2 \cr}
\eqno(2.4Ba)
$$
with the restriction
$$
\alpha-pq\neq 0
\eqno(2.4Bb)
$$
and with the normal ordering condition (1.12)
$$
\alpha\beta^2+1\neq 0
\eqno(2.4Bc)
$$
\vskip 0.5 true cm
\item{Case C}
$$\eqalign{
G*X&=p X*G+X*A   \cr
A*X&=-p X*A+\alpha X*G \cr
G*A&=- A*G+\beta A^2+(\alpha\beta-2p) G^2 \cr}
\eqno(2.4Ca)
$$
with the restrictions
$$\eqalign{
\beta p+1&\neq 0\cr
\alpha +p^2&\neq 0\cr}
\eqno(2.4Cb)
$$
and with the normal ordering condition (1.12)
$$
\beta(\alpha\beta -2p)\neq 1
\eqno(2.4Cc)
$$
\vskip 0.5 true cm
\item{Case D}
$$\eqalign{
G*X&=p X*G+X*A   \cr
A*X&=-p X*A-(2p^2+p\gamma) X*G \cr
G*A&=- A*G-{1\over p} A^2+\gamma G^2 \cr}
\eqno(2.4Da)
$$
with the restriction
$$
p+\gamma\neq 0
\eqno(2.4Db)
$$
while the normal ordering condition (1.12) coincides here with (2.8b).
\vskip 0.5 true cm
\par Obviously, the situation is somewhat complicated by the fact
that for each $i$,
the solution can be chosen arbitrarily to belong to one of the four
cases $(A,B,C,D)$ above.
\vskip 0.5 true cm
The usual Heisenberg algebra belongs to class B.
\vskip 0.5 true cm
\par It may be interesting to ask the question whether, by a suitable
linear change of basic operators, the set of quommutators, which is
written in (2.3) with underlying physical motivations, cannot be
brought in a simpler canonical position. Let us first stress that $X$
plays a special role and that we are thus restricted to consider linear
combinations of $G$ and $A$ only. Let $G'$ (or $A'$) be given by
$$
G'=vG+wA
\eqno(2.5)
$$
It will fulfill the equation
$$
G'X=p'XG'
\eqno(2.6)
$$
provided $p'$ and $\lambda={w\over v}$ satisfy
 the equations
$$\eqalign{
p'^2-(p+q)p'+(pq-\alpha)=&0\cr
\alpha\lambda^2+(p-q)\lambda -1=&0\cr}
\eqno(2.7)
$$
\par If there are two different roots, say $p'$ and $q'$, to (2.7) i.e. if the
discriminant $\Delta$ which is the same for both
equations is non zero
$$
\Delta\equiv (p-q)^2+4\alpha\neq 0
\eqno(2.8)
$$
we see that (2.3) can be brought to the canonical form
$$\eqalignno{
G'*X&=p' X*G'&(2.9a)  \cr
A'*X&=q' X*A'&(2.9b)\cr
s'G'*A'&=r' A'*G'+\beta' A'^2+\gamma' G'^2 &(2.9c)\cr}
$$
where the last equation is the most general quadratic combination of
na\"\i ve zero degree. If we suppose that $s'$ is non zero, it can be
renormalised to 1.
\par If $\alpha=pq$, i.e. case $A$ above, one of roots of (2.7), $q'$ say,
is zero.
\par The associativity requirements for (2.9) are (provided that we suppose
that $s'$ and $r'$ are not both zero and remembering that $q'\neq p'$)
$$\eqalignno{
p'\gamma'&=0       &(2.9d)\cr
q'\beta'&=0        &(2.9e) \cr}
$$
\par Since at least one of the roots is non-zero, say $p'$, (2.9d) implies
that
$$\gamma'=0
\eqno(2.9f)
$$
and then (2.9e) leads to two cases
$$
Case\  A'\ \ :\ q'=0,\ \gamma'=0
\eqno(2.9g)
$$
$$
Case\  B'\ \ :\ \beta'=0,\ \gamma'=0
\eqno(2.9h)
$$
\par When the two roots of (2.7) are equal (i.e. if $\Delta$ of (2.8) is zero
and $p'=(p+q)/2$), the system can
be brought to the form
$$\eqalignno{
GX&=p'X*G+XA'&(2.10a)  \cr
A'*X&=p' X*A'&(2.10b)\cr
s'G*A'&=r' A'*G+\beta' A'^2+\gamma' G^2 &(2.10c)\cr}
$$
where
$$\eqalignno{
A'&=A+{p-q\over 2}G &(2.10d) \cr
p'&={p+q\over 2}    &(2.10e) \cr
s'&=1-{\beta(q-p)\over 2}   &(2.10f) \cr
r'&=r+{\beta(q-p)\over 2}   &(2.10g) \cr
\beta'&=\beta       &(2.10h) \cr
\gamma'&=\gamma+{\beta(p-q)^2\over 4}+{(q-p)(r-1)\over 2} &(2.10i) \cr}
$$
The form of (2.10a), (2.10b) is somewhat similar to that of the well known
deformation of the Heisenberg commutation relations of Biedenharn$^{[10]}$
and Macfarlane$^{[11]}$, with the formal identification of $a^{\dag}$ with $X$,
 $a^{\dag}a=N$ with $G$ and $q^{-N}$ with $A$. Of course in their scheme, the
operators $A$ and $G$ are not independent, as they are for us.

\par If $s'\neq 0$ it can be renormalized to 1 and the associativity
requirements for (2.10) are
$$\eqalignno{
\gamma'&=0       &(2.10j)\cr
p'(s'-r')&=0        &(2.10k) \cr}
$$
leading to two cases again.
$$
Case\  C'\ \ :\ p'=0,\ \gamma'=0
\eqno(2.10l)
$$
$$
Case\  D'\ \ :\ r'=s',\ \gamma'=0
\eqno(2.10m)
$$
\par If $s'=0$, $r'$ can be renormalized to -1 and the associativity
requirements lead to
$$\eqalignno{
&p'\gamma'=0 &(2.10n)\cr
&p'-\gamma'=0 &(2.10o)\cr}
$$
\par This leads to the rather uninteresting case
$$
Case\  E'\ \ :\ p'=0,\ \gamma'=0
\eqno(2.10p)
$$

\par It is obvious that the associativity conditions were easier to write
in the new basis but we first presented the results in
the old one as it is more physical. The change of basis facilitates, in
certain cases, the
discussion of representations.
\par Let us also note that the conditions (2.2) are exactly those which
allow the above redefinitions in terms of $G'$ and $A'$ in a coherent way
between different indices.

\vskip 0.5 true cm
{\bf{III. Symmetric q-derivatives.}}
\vskip 0.5 true cm
\par We now present a particular case of associative operators which are
suited to define a symmetric q-derivative.
\par Let, as above, the $X_i$ be a set of quantum variables. For a
function $f(x_i)$ of one variable alone, let us define
$$\eqalignno{
(X_if)(x_i)&=x_if(x_i)  &(3.1a)\cr
(D_if)(x_i)&={f(p_ix_i)-f(x_i/p_i)\over x_i(p_i-1/p_i)}&(3.1b)\cr}
$$
as the symmetrical q-derivative.
\par As a consequence the operator $G_i$ defined in (1.3) has the
following action
$$
(G_if)(x_i)={f(p_ix_i)-f(x_i/p_i)\over (p_i-1/p_i)}
\eqno(3.2)
$$
\par The operator $A_i$ defined by
$$
D_i*X_i=p_{ii} X_i*D_i+A_i
\eqno(3.3)
$$
has still some arbitrariness but can also be chosen, in a unique
way, to have a symmetric action
$$
(A_if)(x_i)={f(p_ix_i)+f(x_i/p_i)\over 2}
\eqno(3.4)
$$
\smallskip

\par From these definitions the
quommutation relations within the $i$ family can be deduced
$$\eqalign{
G_i*X_i&=p_{ii} X_i*G_i+X_i*A_i \cr
A_i*X_i&=p_{ii} X_i*A_i+{(p_{ii}^2-1)} X_i*G_i\cr
G_i*A_i&=A_i*G_i \cr}
\eqno(3.5)
$$
where
$$
p_{ii}={p_i+1/p_i \over 2}
\eqno(3.6)
$$
\par They correspond to Case B above with $q=p$, $\alpha={p^2-1}$ and
$\beta=0$, a particularly simple case.\par The connection between the operators
coresponding to two different
indices $i$ and $j$ still depend on four a priori independent parameters
$x_{ij}$, $g_{ij}$, $p_{ij}$ and $p_{ji}$.
However the consistency of the quommutator (1.7) when applied to the
function $f\equiv 1$ implies $a_{ij}=1$ i.e. $g_{ij}=1$ since through
(3.4) the action of
$A_i$ on $1$ is $1$ for any $i$.
\par Let us note also that a exponential can be easily defined as a
solution, say for one variable $x$ only, of
$$
D E(x)=E(x)
\eqno(3.7)
$$
For the symmetrical q-derivative it reads
$$
E(x)=\sum_{n=0}^{\infty} a_nx^n
\eqno(3.8)
$$
where
$$\eqalign{
a_0&=1,\ \ \ \ a_1=1 \cr
a_n&={1\over \prod_{k=1}^{n-1}[k]_p}\ \ \ n>1.\cr}
\eqno(3.9)$$
Here
 $$[j]_{q_i}={ q_i\sp j-q_i\sp {-j}\over q_i-q_i^{-1}}.\eqno(3.10)$$
  In order to generalise the q-exponential to many variables it is necessary
to  choose a particular set of values for the parameters $p_{ij}$ etc.
This choice is
$$\eqalign{p_{ij}=q_{ij}=& {(q_i+q_i^{-1})\over2},\ \ \forall j,\cr\alpha_i=&
 {(q_i-q_i^{-1})^2\over4},\cr
r_{ij}=& 1,\ \gamma=0.\cr}\eqno(3.11)$$

It allows $x_ix_j$ to be interchanged; it turns out that
$$(q_j+q_j^{-1})x_ix_j=(q_i+q_i^{-1})x_jx_i\quad\forall i,j\eqno(3.12)$$
may be imposed without affecting the q-exponential. This equation allows all
 monomials in $x_i,x_j,\dots$ to be ordered alphabetically.

The general q-exponential is given by the simultaneous solution of the
equations
$$ D_iE(x_1,x_2,\dots x_N)=E(x_1,x_2,\dots x_N).     \eqno(3.13)$$

 The general term in
 the expression $E(x_1,x_2,x_3)$ for example, is given by
$$\prod\sp{a}_j{1\over[j]_{q_1} }\prod\sp{b}_k{1\over[k]_{q_2}}
\prod\sp{c}_l{1\over[l]_{q_3}}
 {x_1\sp ax_2\sp bx_3\sp c}
\Bigl({2\over[2]_{q_1}}\Bigr)\sp{a(b+c)}\Bigl({2\over[2]_{q_1}}\Bigr)\sp{bc}.
\eqno(3.14)$$
The generalisation to an arbitrary number of variables is obvious.
Thus this choice of parameters allows a simultaneous construction of a
generalised q-exponential, or coherent state, and a set of permutable creation
operators. This last feature is admittedly absent in the work of
Greenberg$^[3]$ and in most of the literature on coherent states by
implication.
\vskip 0.5 true cm
{\bf{IV. General q-derivatives.}}
\vskip 0.5 true cm
\par A slight generalisation of the preceding section can be obtained as
follows in the case of one variable only.
\par Suppose we define the $X$ as before on a function $f(x)$
$$
(Xf)(x)=xf(x)
\eqno(4.1)
$$
and try to construct the action of $G$ as
$$
(Gf)(x)=\sum_{k=1}^M \chi_kf(\lambda_kx)
\eqno(4.2)
$$
Equation (2.3a) is then a simple definition of $A$
$$
(Af)(x)=\sum_{k=1}^M \chi_k(\lambda_k-p)f(\lambda_kx)
\eqno(4.3)
$$
Equation (2.3b) is then a consistency equation which reads
$$
\sum_{k=1}^M \chi_k\left(\lambda_k^2-(p+q)\lambda_k+qp-\alpha\right)
f(\lambda_kx)=0.
\eqno(4.4)
$$
Since the $f(\lambda_kx)$ are independent for sufficiently general choices
of $f(x)$ (4.4) implies that
$$
\lambda_k^2-(p+q)\lambda_k+qp-\alpha
\eqno(4.5)
$$
for every $k$ (equation (2.7) again). But since $p,q$ and $\alpha$ don't depend
on $k$ and since
also a second degree equation has only two solutions, there are at most
two allowed values of $\lambda_k$. Hence $M=2$ and the two $\lambda$'s are
given in terms of the three free parameters $p,q$ and $\alpha$.
\par Conversely, if the two values of $\lambda$ are given
$\lambda_1=\lambda$, $\lambda_2=\mu$, the condition (4.5) gives the
restrictions
$$\eqalignno{
p+q&=\lambda+\mu &(4.6)\cr
pq-\alpha&=\lambda\mu &(4.7)\cr}
$$
So that there are altogether again
three free parameters, say $\lambda,\mu$ and $p$. The remaining ones $q$
and $\alpha$ being fixed
by (4.6) and (4.7).
\par Finally $A$ and $G$ commute
$$
G*A=A*G.
\eqno(4.8)
$$
\par All these equation finally generate a representation of case B with
$\beta=0$ but the other parameters are free. This representation can be
extended in a natural way when there is more than  one variable $x_i$.
Obviously these variables are then
quantum variables and have to fulfill
quommutation relations in agreement with (1.1a).
\par The representations discussed so far are those acting on an infinite
 dimensional function space.
\vskip 0.5 true cm
{\bf{V. Representations with $G$ and $A$ diagonal.}}
\vskip 0.5 true cm
Let us look now for representations such that both $G$ and $A$ are
diagonal and hence commute.
\par Suppose we start with a vector $\mid 0>$, eigenvector of $G$ and $A$ with
eigenvalue $g_0$ and $a_0$ respectively
$$\eqalign{
G\mid 0>&=g_0\mid 0>\cr
A\mid 0>&=a_0\mid 0>\cr}
\eqno(5.1)
$$
Let us define
$$
\mid n>=X^n\mid 0>
\eqno(5.2)
$$
Then
$$\eqalign{
G\mid n>&=g_n\mid n>\cr
A\mid n>&=a_n\mid n>\cr}
\eqno(5.3)
$$
\par With the vector $v_n$ defined by
$$
v_n={\left( {\matrix{g_n\cr
                     a_n\cr}} \right)}
\eqno(5.4)
$$
and the matrix $M$ defined by
$$
M={\left( {\matrix{p & 1\cr
                   \alpha & q\cr}} \right)}
\eqno(5.4)
$$
it is easy to prove that
$$
v_n=Mv_{n-1}=M^n v_0
\eqno(5.5)
$$
\par This obviously depends on the
precise form of the matrix $M$ and of its eigenvalues,
obviously equation (2.7) again.
\par Let us now make the following important remark. Since both $G$ and
$A$ are diagonal, they commute for this representation. But this
commutation is {\bf{not}} a qualgebra relation. What has to hold is
(2.5c) which has still to be satisfied
at every stage of the procedure. It reads in the general case
$$
C_n\equiv \beta a_n^2+\gamma g_n^2+(r-1)a_ng_n=0
\eqno(5.6)
$$
\par More precisely if, say $a_n$ and $g_n$ are chosen in such a way
that $C_n=0$ there is a condition on the free parameters to garantee that
$C_{n+1}$ be zero. Either
$$\eqalign{
&p^2\beta\gamma-pqr^2+2pqr+2pq\beta\gamma-pq-pr\gamma+pr\alpha\beta+p\gamma
-p\alpha\beta+q^2\beta\gamma\cr
&+qr\gamma-qr\alpha\beta-q\gamma+q\alpha\beta
+\gamma^2-2\alpha\beta\gamma+\alpha^2\beta^2=0\cr}
\eqno(5.7a)
$$
or
$$\eqalign{
&p^2\gamma\beta-2pq\beta\gamma-pr\gamma+pr\alpha\beta+p\gamma
-p\beta\alpha+q^2\beta\gamma\cr
&+qr\gamma-qr\alpha\beta-q\gamma+q\alpha\beta
-r^2\alpha+2r\alpha+\gamma^2+2\alpha\beta\gamma+\alpha^2\beta^2-\alpha=0
\cr}
\eqno(5.7b)
$$
\par It is amusing to note that the product of the two expressions is
always identicallly zero in the four allowed cases $A-D$. This fact shows
that this representation always exists. Let us stress again that the
fact that $G$ and $A$
commute is a simple artefact of the representation. It is analogous for
example to the fact that, if the Pauli matrices are 2-dimensional
representations of $SU(2)$, the fact that $\sigma_1\sigma_2 =i\sigma_3$
is a simple artefact which has nothing to do with the basic commutation
relations of the $SU(2)$ algebra.

\vskip 0.5 true cm
{\bf{VI. Representations where $X$ has an eigenvector.}}
\vskip 0.5 true cm
\par In the preceding paragraph we have chosen to present the case where
both $G$ and $A$ are diagonal, as they are simple. However we believe that
the physically interesting representations rather correspond to cases where
$X$ has  an eigenvector $\mid 0>$ with eigenvalue $x_0$
$$
X\mid 0>=x_0\mid 0>
\eqno(6.1)
$$
It is now more convenient to re-introduce the linear combinations in terms of
which the quommutation relations simplify to (2.9).\par An infinite set of
states $\mid k,l>,\ k=1,\ldots,l;\ l=1,\ldots$ can
then be constructed through
$$
\mid k,l>=A'^kG'^l\mid 0> ,\ \ \ \ 0\leq k\leq l
\eqno(6.2)
$$
 The action of the operators in this infinite dimensional space in the case
can then be found easily through the
equations
$$\eqalign{
X\mid k,l>&=x_0 q'^{-k}p'^-l\mid k,l>\cr
A'\mid k,l>&=\mid k+1,k>\cr
G'\mid k,l>&=\mid k,l+1>\cr}
\eqno(6.3)$$
When  $G'$ and $A'$  do not commute then the last of these equations requires
 modification.
Using (2.9c), applied to the states $\mid l,0>$, $\mid l+1,1>$,
$$\eqalign{
G'\mid 1,k>=&r\mid 1,k+1>+\beta\mid 2,k> +\gamma\mid 0,k+2>\cr
G'\mid 2,k>={1\over (1-\beta\gamma)}
              &{\big(} \beta(1+r)\mid 3,k> +\gamma^2(1+r) \mid 0,k+3>\cr
         & +r\gamma(1+r)\mid 1,k+2>+r(r+\beta\gamma)\mid 2,k+1>{\big)} \cr}
\eqno(6.4)$$
The last of these equations also follows from (1.13) applied to $\mid l,0>$.
The general result will follow upon iteration.
\vskip 10pt
\vskip 0.5 true cm
{\bf{VII. Finite dimensional representations.}}
\vskip 0.5 true cm
\par Finite dimensional representations of operators are always useful
to consider.
We restrict ourselves to situations where $p\neq 0$, $q\neq 0$ and
$r\neq 0$.
\vskip 0.5 true cm
{\bf{One dimensional representations.}}
\vskip 0.5 true cm
\par First, the one dimensional representations are obviously trivial but
we present them in order to be able to make the remark following (7.2) below.
\par When $X$ is represented by $1$ by na\"\i ve rescaling and $G=1$ also.
\par Then
$$\eqalign{
X&=1\cr
G&=1\cr
A&=(1-p)\cr}
\eqno(7.1a)
$$
We can apparently solve directly
for the
parameters of (2.3), without
going through the four cases (A-D) one by one.
\par The parameters are restricted
by the two relations
$$\eqalignno{
\alpha &=(1-p)(1-q)&(7.1b)\cr
\gamma &=(1-p)(1-r-\beta +p\beta)&(7.1c) \cr}
$$
\par When $X=0$ then
$$\eqalign{
X&=0\cr
G&=g\cr
A&=a\cr}
\eqno(7.2a)
$$
and the numbers $g$ and $a$ must satisfy
$$
(1-r)ga-\beta a^2-\gamma g^2=0
\eqno(7.2b)
$$
\par It may appear strange at first sight that the representation (7.1)
(or (7.2))
looks more general than any of the cases (A--D) above. The
justification for this fact is as follows. To derive the conditions of
associativity we have supposed, rightly, that, once put in normal
order, any product of the starting operators $X,G,A$, for up to three
operators in the product, is linearly independent of any other. This is
clearly not true for $X=G=1$. Hence there are apparently more solutions
to the associativity requirements. These extra solutions should
obviously be rejected as they are not bona fide representations of the
abstract quommutators.
\vskip 0.5 true cm
{\bf{Two dimensional representations.}}
\vskip 0.5 true cm
\par Henceforth we will restrict ourselves to representations where not
all the operators are represented by diagonal matrices, i.e. irreducible
representations.
\par By performing a general change of basis in the two-dimensional
space upon which the operators act and by using a suitable rescaling of
the naive length $[L]$ unit, it is always  to bring the
operator $X$ in one of the following well-known canonical positions :
\item{a)} $X=1_2$ is the 2-dimensional unit matrix
\item{b)} $X =$ diag$(1,x)$ where $x\neq 0$ and $x\neq 1$
\item{c)} $X =$ diag$(1,0)$
\item{d)} $X =\sigma_+$ where $\sigma_+$ is the Pauli matrix with only
non-zero element $\sigma_+(1,2)=1$
\item{e)} $X =1_2+x\sigma_+$ where $x\neq0$.
\par The full discussion is rather rich. Indeed, the
allowed representations depend often on particular and more detailed
relations between the parameters than those which define the four cases
(A--D) above. We have thus chosen not to present them though some of
them are quite interesting.
\vskip 0.5 true cm
{\bf{ Finite representations with $G$ and $A$ diagonal.}}
\vskip 0.5 true cm
\par Using the results of section V, finite dimensional representations
of the qualgebras can be constructed.  Indeed, if we suppose that there
exists a positive integer $P$ such that (see (5.2))
$$
\mid P>=\mid 0>
\eqno(7.3)
$$
the space on which the representation acts becomes $P-$dimensional. In
order to reproduce the same eigenvalues for $G$ and $A$, one must have
$v_P=v_0$. Consequently one needs (see (5.5))
$$
M^P=1
\eqno(7.4)
$$
\par For this to be the case, the two eigenvalues of $M$ i.e. $p'$ and
$q'$ have to be $P-$roots of unity. We find once again
the occurrence of
the important ``roots of unity'' which play a central role in qualgebras.
\par In the transformed basis where $M$ is diagonalized, X is
essentially a matrix of cyclic permutation.
\par Other representations can be constructed from these by using
suitable direct sums or direct products of representations. An example
using direct
products, usually equivalent to one of the general type constructed above, or
reducible to direct sum of them
 is as follows :
If $a$ and $b$ in are
prime numbers, then a representation by $ab\times ab$ dimensional matrices in
the changed basis is
given by
$$\eqalign
{ G=&{\rm diag}\{ 1,p',p'^2,\dots p'^{a-1}\}\otimes
{\rm diag}\{ 1,1,1,\dots 1\}\cr
A=&{\rm diag}\{ 1,1,\dots 1\}\otimes{\rm diag}\{ 1,q',q'^2\dots q'^{b-1}\}\cr
X=&P_{a\times a}\otimes P_{b\times b}\cr}
\eqno(7.5)
$$
Here $p'^{a}=1,\ q'^{b}=1$ and $P_{a\times a}$ and  $P_{b\times b}$  are
matrix representations of cyclic permutations of $a$ and $b$ objects
 respectively. Here again $G$ and $A$ commute but, as explained above
this is a simple artefact of the representation of a more general qualgebra.
\vskip 0.5 true cm
{\bf{VIII. Conclusion.}}
\vskip 0.5 true cm
\par Using a natural set of a priori quommutation relations
between quantum variables $X_i$, quantum derivatives $D_i$ or better
the related quantum dilatation operators $G_i=X_i*D_i$ we have been led to
introduce  the
corresponding neutral operators $A_i$, in order to give an algebraic
realisation of symmetric q-differentiation. We have
outlined all the possible choices allowed by the associativity
requirements (or braiding relations) within our basic restriction (1.1a), that
the transposition of two quantum variables does not intoduce any other
operators. We have shown that, within a given
$i$ (the $i$ family) and taking into account the normal ordering
condition, there are four disconnected cases, (2.4A-D). The relations
beween two families $i$ and $j$ depend, due to (2.2), on four arbitrary
parameters
$x_{ij}=1/x_{ji}$, $g_{ij}=1/g_{ji}$,
$p_{ij}$ and $p_{ji}$.
 We have presented some representations
of our abstract qualgebras with one $X$, one $G$ and one $A$ only
both in finite and infinite
dimensional spaces. Having constructed all
2-dimensional representations, we have realized that the space of
representation is apparently very rich.
\par A particular example of a representation for these operators and
their action has been shown explicitly for what we have called
symmetric or general quantum derivatives.
It allows $x_ix_j$ to be interchanged and at the same time permits the
construction of a multivariable  q-exponential; explicitly the result is
obtained that
$$(q_j+q_j^{-1})x_ix_j=(q_i+q_i^{-1})x_jx_i\quad\forall i,j\eqno(8.2)$$
may be imposed without affecting the q-exponential. This may have some bearing
upon attempts to apply quantum groups to quantum optics. We hope to
 elaborate on those points in the near future.

\vfill\eject
\centerline{References}
\vskip 1 true cm
\item{[1]} Woronowicz SL, {\it Comm. Math. Phys.} {\bf 122} (1989) 125.
\item{[2]} Fairlie  D.B. and  Nuyts J., Neutral and charged quommutators
with and without symmetries, Zeitschrift f\"ur Physik 56(1992)237.
\item{[3]} Greenberg O.W., {\it Phys. Rev. Lett.}\ {\bf
64}\ (1990), 705.
\item{[4]}Bogoliubov N.M. and Bullough, R. K. {\it J. Phys A} {\bf A25} (1992)
4057.
\item{[5]}Pusz, W. and Woronowicz, S.L.,
{\it{Rep.Math.Phys.}}, {\bf{27}} (1989) 231.
\item{[6]}Vokos, S.P.,
Zumino, B. and Wess, J., {\it{Z.Phys. C}}, {\bf{48}} (1990) 317.
\item{[7]}Wess J. and Zumino B. {\it Nuclear Physics B Supplements} {\bf 18B}
(1990) 302.
\item{[8]} Zumino B., {\it Mod. Phys. Lett.} {\bf A13} (1991) 1225.
\item{[9]} Fairlie D.B. and  Zachos C.K., {\it Phys.Lett.}
{\bf 256B} (1991) 43.
\item{[10]}   Biedenharn, L.C. {\it J.Phys. A} {\bf A22} (1989) L873.
\item{[11]}  Macfarlane, A.J  {\it J.Phys. A} {\bf A22} (1989) 4581.
\vfill\eject
 \end